# Node, Node-Link, and Node-Link-Group Diagrams: An Evaluation

Bahador Saket, Paolo Simonetto, Stephen Kobourov and Katy Borner
Computer Science Department, University of Arizona
Department of Information and Library Science, Indiana University

## Abstract

Effectively showing the relationships between objects in a dataset is one of the main tasks in information visualization. Typically there is a well-defined notion of distance between pairs of objects, and traditional approaches such as principal component analysis or multi-dimensional scaling are used to place the objects as points in 2D space, so that similar objects are close to each other. In another typical setting, the dataset is visualized as a network graph, where related nodes are connected by links. More recently, datasets are also visualized as maps, where in addition to nodes and links, there is an explicit representation of groups and clusters. We consider these three *Techniques*, characterized by a progressive increase of the amount of encoded information: node diagrams, node-link diagrams and node-link-group diagrams. We assess these three types of diagrams with a controlled experiment that covers nine different tasks falling broadly in three categories: node-based tasks, network-based tasks and group-based tasks. Our findings indicate that adding links, or links and group representations, does not negatively impact performance (time and accuracy) of node-based tasks. Similarly, adding group representations does not negatively impact the performance of network-based tasks. Node-link-group diagrams outperform the others on group-based tasks. These conclusions contradict results in other studies, in similar but subtly different settings. Taken together, however, such results can have significant implications for the design of standard and domain specific visualizations tools.

## 1 Introduction

Information spatialization combines techniques from cartography, statistics, and perception psychology to visualize non-spatial data. Objects in non-spatial data do not have a strong connection with a position in space, either because they are purely abstract, or because they do not have a real spatial dimension or an established convention about their placement.

Spatialization methods place these objects in 2D or 3D space so that the first law of geography (closer things are more similar) [32] is respected. Since this requires a predefined concept of similarity, the data to be spatialized often comes with, or is subsequently divided in, clusters of similar objects. Therefore, the results often resemble geographical maps, with groups of related nodes as countries.

*Scatter plots* are a very traditional spatialization, frequently used in the natural sciences to find patterns and groups in empirical bivariate data. Scatter plots date back to as early as 1833, when the mathematician and astronomer J. Herschel studied the relationship between magnitude and spectral classes of stars. According to Tufte [51] "the relational graphic—in its barest form, the scatterplot and its variants—is the greatest of all graphical designs." With the success of dimensionality reduction techniques such as principal component analysis and multi-dimensional scaling, scatter plots and point cloud visualizations are a powerful tool in the statistical visualization toolbox.

*Node-link diagrams* date back to the 18th century and the "seven bridges of Königsberg" problem, modeled by L. Euler with nodes (for the different parts of the city) and links (for the bridges between them). Such relational datasets are typically characterized by a set of objects (e.g., webpages) and relationships between them (e.g., links between pages). Graph drawing algorithms, or network layout methods, are another standard tool in the visualization toolbox in many fields from software engineering, bioinformatics, to social network analysis.

*Map-based visualizations* are among the oldest visualizations [8, 7], and placing imagined places on imagined maps has a long history, e.g., the 1930s Map of Middle Earth by Tolkien. A more recent example is xkcd's Map of Online Communities [34]. While most such maps are generated in an ad hoc manner and are not strictly based on underlying data, they are often very visually appealing. The map metaphor is a particularly popular approach in the context of text visualization [46, 53], and recently a number of fully automated tools were developed to generate such map-like visualizations for non-spatial data.

In this paper we consider these three visualizations, commonly employed in spatialization, which for the purpose of uniformity we call *node diagrams* (N diagrams), *node-link diagrams* (NL diagrams) and *node-link-groups diagrams* (NLG diagrams). Each of these diagrams extends the previous one by making more explicit a characteristic of the input data.

In N diagrams, a set of objects is depicted as points in a two or three dimensional space; see Figure 1a. Clusters are typically depicted by painting each node with a color that is unique for each group. Such diagrams are very common in natural sciences and are generated by principal component analysis (PCA) [27] or multi-dimensional scaling (MDS) [28]. Such visualizations are often referred to as scatter plots, scatter diagrams, and point clouds [36].

In NL diagrams, the visualization is enriched with connections that make explicit a close relation between two elements; see Figure 1b. As before, colors are typically used to indicate group membership. Node-link diagrams are often referred to as graphs drawings, or network layouts and are the standard way of representing relational data [6, 20].

In NLG diagrams, the visualization is further enriched by enclosing the elements that belong to the same set into a region; see Figure 1c. This is the output of several recent InfoVis techniques

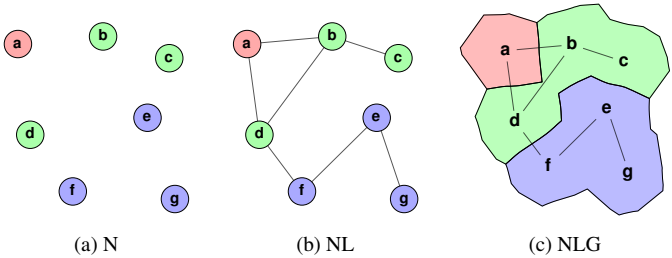

(a) N  (b) NL  (c) NLG

Figure 1: Examples of diagrams considered in this study.

which visualize sets, groups, and clusters [12, 21, 31, 14].

N diagrams offer an effective way to show clear partitions of the data. However, PCA, MDS and similar techniques might obscure some details. By explicitly drawing a link between closely related objects, NL diagrams can show a related pair of objects, even when the objects are not nearby. Enclosing the elements of the same group in a region in NLG diagrams makes grouping explicit, provides a high-level structural overview, and alleviates potential problems with color ambiguities.

In this paper, we consider the effectiveness of these three types of visualizations (*Techniques*) on node-based tasks, network-based tasks and group-based tasks, with a controlled experiment.

## 2 Related Work

There are various approaches to data spatialization in different disciplines: scatterplots in statistics and the natural sciences [27, 28, 18], abstract maps in cartography [45, 46, 48, 47] and in visual arts [7, 22], node-link diagrams in graph drawing [4, 9] and Euler/Venn diagrams in set visualization [40, 52, 19, 44].

A great deal of related work evaluates the general concepts of spatialization and specific spatialization techniques.

**N diagrams:** The readability of node diagrams has been studied for nodes and groups of nodes. There is evidence that the distance between pairs of nodes is related to the perceived similarity between them [17], but it known that this can be significantly altered by other factors, including boundaries used to group nodes [16]. The relative position and arrangement of nodes also influence the perceived importance of the nodes. Central nodes are generally perceived as more important, while regular node arrangements, such as placing the nodes around a circle, tend to suggest that the nodes involved are equally important [30, 13]. Node spacing is particularly important in the perceived clustering, as changes in node proximity induce the users to detect different number of clusters and of nodes that act as bridges between one group and another [30]. Finally, several studies have considered how to depict the group boundaries, defining patterns that should and should not be present, as well as evaluating their impact on the diagram comprehension [49, 5, 41].

**NL diagrams:** The readability of graph and network layouts has also been studied. In graphs, the placement of the nodes and links can result it desirable (e.g., display of symmetries) or undesirable results (e.g., edge crossings). The impact of such aesthetic criteria has been evaluated [37, 39], showing that some have a significant impact on readability (*e.g.* the number of edge crossings), while others have statistically insignificant effects. Metrics have been developed to formally evaluate some of these aesthetic criteria [38]. In the latest study, Alper et al. [2] compared node-link diagrams with matrix representations, using a controlled experiment to assess which representation best support weighted graph comparison tasks.

**NLG diagrams:** There is less work on evaluating node-link-group diagrams, as these are fairly new. Very recently, Jianu et al. [26] evaluated four techniques for displaying group or cluster information overlaid on node-link diagrams: node coloring, GMap [21], BubbleSets [12], and LineSets [3]. The focus of the study is to match specific tasks to specific visualizations. BubbleSets were found to outperform the other visualizations in tasks that involve group perception and understanding.

Tory et al. [50] compared the performance of search and point-estimation tasks on N diagrams and 2D/3D landscapes, that closely resemble NLG diagrams, but do not have edges/links. Their results show that N diagrams outperform landscapes, and that using the third dimension is detrimental for these drawings. However, this does not directly answer the questions posed in our paper for a couple of reasons. First, in [50] the focus is on points and their metric values, whereas we also study the relations between the objects and between groups of objects. Second, groups are identified by splitting the range of a metric into different intervals and creating groups that collect all nodes in that interval. Thus colors are not only used to identify the groups, but also to provide quantitative information about the value of the metric. It is therefore necessary to find a balance between two conflicting needs: providing a color scale that facilitates the estimation of the metric (e.g., increasing color saturation) versus providing a color scale that provides good distinctions between the groups (e.g., rainbow scale). We do not have such a conflict in our setting.

### 2.1 Group Visualization

The most related prior work is that of Jianu et al. [26]. There are several factors that impact the conclusions in that study: contiguity, clutter, and features. We briefly discuss these below:

**Contiguity:** BubbleSets and LineSets produce contiguous regions, whereas GMap produces fragmented regions. As pointed out in [26], for some tasks such as "asking users to see whether two nodes are located in the same group or not", the user performance highly depends on how the two nodes are selected. If the two highlighted nodes are located in the same fragment, then user performance may not change in both BubbleSet and GMap, while if both highlighted nodes are in spatially scattered fragments that belong to the same group, then GMap cannot compete with BubbleSets. We avoid this problem by using only contiguous regions in our NLG diagrams.

**Clutter:** There are different types of visual clutter introduced by the visualizations studied in [26] which affect the results. GMap introduces clutter by displaying group labels over distinctive sets. As pointed out by the authors, such group labels in GMap caused invalid results in some of the tasks. For example, the task of "Estimating the degree of a highlighted node", is impossible when the group label is located on the top of neighbors of the node. Similarly, BubbleSets introduces clutter in areas where multiple groups overlap. We avoid this problem by eliminating all types of clutter in our three visualizations.

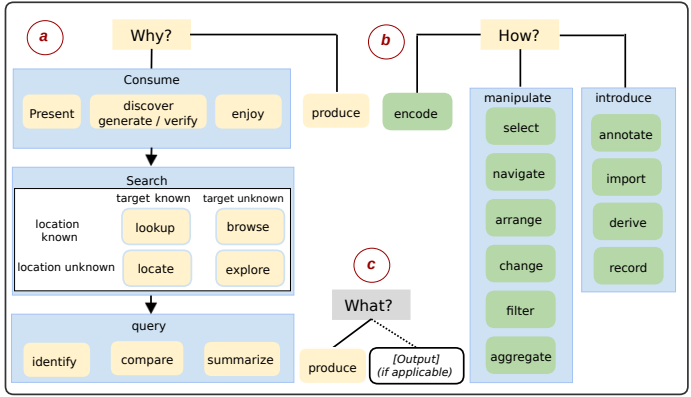

Figure 2: Multi-level typology of abstract visualization tasks. The typology spans WHY, HOW and WHAT. Figure from [10] used with permission.

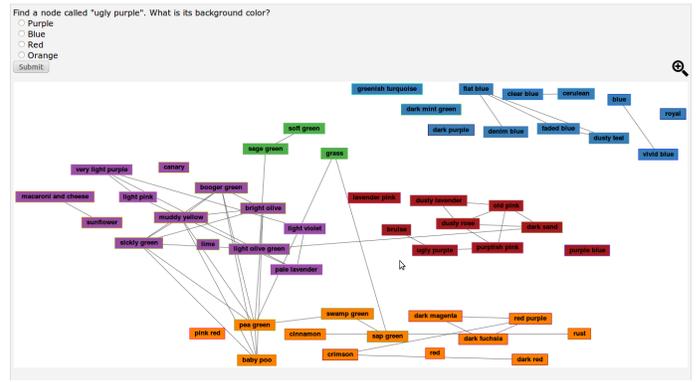

Figure 3: The software guides the participant through the experiment by providing the task instruction and collecting time and accuracy.

**Features:** There are several features of the input data that certainly have an impact on the results (e.g., the number of objects, the density of the network, etc.) In [26] only one dataset with fixed *Size* and *Density* is used. We use several datasets and vary *Size* and *Density* as advocated by [43, 25].

In summary, many earlier studies successfully assess either different aspects of a particular type of visualization, or different types of visualizations. But several big and important questions remain open. We are particularly interested in the effect of adding more information (from nodes only to nodes and links, from nodes and links to nodes and links and groups) on various tasks. Is it harder to perform node-based tasks in an NL or NLG diagrams (compared with an N diagram)? Is it harder to perform network tasks in a NLG diagram (compared with an NL diagram). What is the impact of *Size* and *Density* on the different types of diagrams?

### 2.2 Task Taxonomies

The results of some of the earlier evaluation studies are difficult to compare. Seemingly non-influential decisions, such as the choice or phrasing of the tasks, may have a significant impact on the results. In an attempt to mitigate this problem, visual data analysis tasks are organized and categorized in taxonomies and the literature is rich in such taxonomies. Brehmer and Munzner [10] organized the vast previous work highlighting advantages and disadvantages. They point out as the major shortcoming of most approaches, the lack of a global view of the task: high-level categories often ignore how the tasks are performed, while low level categories often ignore why the tasks are performed. In order to close this gap, they develop a multi-level typology that helps create a complete description of a task.

This multi-level typology encompasses three main questions: WHY, HOW and WHAT. The WHY part of the typology allows us to describe why a task is performed, includes multiple levels of specificity, and a narrowing of scope from high-level (consume vs. produce) to mid-level (search) to low-level (query); see Fig. 2a. The HOW part of the typology allows us to describe how a task is performed, and this part includes three classes of methods: those for encoding data, those for manipulating existing elements in a visualization, and those for introducing new elements into a visualization; see Fig. 2b. Finally, the WHAT part of the typology allows us to describe what are the inputs and outputs for a given task; see Fig. 2c. This definition is purely abstract and enables the translation of any type of relevant task into the why/how/what framework, making it clear and almost ready for implementation.

The work of Brehmer and Munzner, however, is not meant to replace model-oriented taxonomies, but rather to "encompass and complement these specific classification systems". Instead, they provide the tools to put these low level tasks in context, guiding the evaluation designer in providing information, such as user expertise and motivation. We make extensive use of this multi-level typology in our study.

## 3 Controlled Study

In this study we investigates the effectiveness (accuracy, task completion time) of the described N, NL and NLG diagrams. Our aim is to assess how the three *Techniques* scale with changing *Sizes* (changing number of nodes) and *Densities* (changing number of links/edges) across different comparison tasks, to inform designs that would utilize these *Techniques*.

The total number of questions in the main experiment is $\#Questions = \#Sizes \times \#Densities \times \#Tasks$. In order to make the controlled experiment of reasonable length, we need to limit the number of different values of these factors. For *Sizes* and *Densities*, we use three different values, as the minimum requirement needed to provide an estimate of the variation trend. We select values in a geometric progression in order to provide a larger range of considered values. These values are referred to as $N$, $2N$ and $4N$ for *Sizes*, and $L$, $2L$ and $4L$ for *Densities*.

For *Tasks*, we use nine tasks in total, with three tasks per category. This provides the minimum requirement to see variations within a task category.

Table 1: List of tasks used in the evaluation.

| | WHY | WHAT | HOW |
|---|---|---|---|
| **Node-based Tasks** | | | |
| **T1**. Given node "X", what is its background color? | The purpose of the task is to discover the background color of node X. Search target is given (node X) but the node location is not given. Once the participant finds the node they need to identify its background color.<br>(DISCOVER+ LOCATE+ IDENTIFY) | The input for the task is name of a node. The output is the background color of the node.<br>`Input`: Node X<br>`Output`: Background color | Participants need to be able to tell the background color of the node.<br>(DERIVE + SELECT) |
| **T2**. Find all nodes which start with specific alphabet letter in the specific group. | The purpose of the task is to provide list of all nodes which start with a specific alphabet letter (e.g., Z/z). Search target is known since participants need to search for all nodes starting with the specific letter in the specific group. Location of nodes is not given. Finally participants need to produce a list of matching nodes.<br>(DISCOVER + LOCATE + SUMMARIZE) | The input for the task is a letter. The output is the list of nodes which start with that specific letter.<br>`Input`: Specific alphabet letter<br>`Output`: List of nodes | Participants need to be able to identify the nodes with specific alphabet letter.<br>(SELECT) |
| **T3**. What is the number of nodes in a specific group? | The purpose of the task is to count the nodes in a given group. The targets are nodes in the group and the location is the whole group. Thus, both targets and location are known.<br>Participants need to identify the nodes in the group and count them.<br>(DISCOVER + LOOK UP + SUMMARIZE) | The input for the task is a specific group and nodes within it. The output is the number of nodes in that group.<br>`Input`: Nodes in a group<br>`Output`: Number of nodes | Participants need to count number of nodes in the group.<br>(DERIVE) |
| **Network-based Tasks** | | | |
| **T4**. Given nodes X and Y, find the shortest path between them. | The purpose of the task is to identify the shortest past between two given nodes. Targets are given (nodes X and Y) but their location is not given. After finding the nodes participants need to identify paths between nodes X and Y, compare these paths, and find the shortest one.<br>(DISCOVER + LOCATE + COMPARE) | The input for the task are two nodes. The output is the number of links along the shortest path between them.<br>`Input` : Node X and Y<br>`Output`: Shortest path length | Participants need to count number of links in each path between X and Y and identify which path has the fewest number of links.<br>(DERIVE + COMPARE + SELECT) |
| **T5**. Find the set of nodes adjacent to a given node. | Target is given (e.g., node X) but the location of the target is not given. Participants need to produce a list of nodes directly connected to the given node.<br>(DISCOVER+ LOCATE + SUMMARIZE) | The input is a specific node. The output is list of nodes adjacent to the given node.<br>`Input`: Specific Node<br>`Output`: List of nodes | Participants need to distinguish nodes directly connected to the given node.<br>(SELECT) |
| **T6**. Find a node with highest degree. | The purpose of the task is to discover a node with the most incident links. Target is unknown and location is unknown. The participants need to compare nodes with high degree and decide which has the highest degree.<br>(DISCOVER + EXPLORE + SUMMARIZE) | The input is the whole diagram and the output is a node with highest degree.<br>`Input`: Whole diagram<br>`Output`: Specific node | The participant needs to count (and/or estimate) the number of links incident to each node and keep track of the largest ones.<br>(DERIVE + SELECT) |
| **Group-based Tasks** | | | |
| **T7**. Given nodes X and Y, decide whether these two nodes belong to the same group. | The purpose of the task is to discover whether two given nodes belong to the same group. The two nodes are given so the targets are given but their location is not given. Once the participants find both nodes, they need to identify whether they are in the same group or not.<br>(DISCOVER + LOCATE + IDENTIFY) | The input are nodes X and Y. The output is Yes if the two nodes are located in the same group, and No otherwise.<br>`Input`: Nodes X and Y<br>`Output`: Yes/No | Participants need to distinguish whether the two nodes are located in the same group.<br>(SELECT) |
| **T8**. Find the path X—Y—Z; are nodes X and Z in the same group? | The purpose of the task is to discover whether two nodes connected by a path are in the same group. The targets are known nodes X, Y and Z. The location of the three nodes is unknown. Once the participants finds the nodes, they need to determine whether they are in the same group or not.<br>(DISCOVER + LOOK UP + IDENTIFY) | The input for the task are nodes X, Y, and Z. The output is Yes if two nodes are in the same group and No otherwise<br>`Input`: Nodes X, Y and Z<br>`Output`: Yes/No | Participants need to distinguish whether two nodes are located in the same group.<br>(SELECT) |
| **T9**. Given a group X, find the group neighbors of group X. | The purpose of the task is to discover groups that are adjacent to group X. The targets are known (X is specified). The location of the group X is not mentioned so the location is unknown. The participants need to produce of list of groups which have common boundaries with the given group X.<br>(DISCOVER + EXPLORE + SUMMARIZE) | The input is a specific group. The output is list of groups that are neighbors of the given group.<br>`Input`: Group X<br>`Output`: List of groups | The participants need to identify group X and the groups which have common boundary with group X.<br>(SELECT) |

## 3.1 Tasks

We first considered user interactions with visualization systems such as BubbleSets [12], LineSets [3], and GMap [21]. We also considered existing task taxonomies for graph visualization [29], and interviewing several experts in the field. The result was a list of over 80 different tasks, which we divided into three categories according to the information required to solve them.

- **Node-Based Tasks**: Tasks in this category can be performed by considering only nodes, so that no other information is required. *For example: Given node "X", what is its background color?*

- **Network-Based Tasks**: Tasks in this category can be performed by considering only nodes and links. *For example: Find a node with the highest degree.*

- **Group-based Tasks**: Tasks in this category can be performed by considering nodes, links, and groups. *For example: Given a group X, find all groups neighboring group X.*

We looked for simple tasks that can be performed in a reasonable amount of time and validated them using Brehmer and Munzers multilevel typology. Most of the tasks in the first two categories are listed under "Attribute-Based Tasks" and "Topology-Based Tasks" in the work done by Lee et al. [29]. Most of the tasks in the third category are "Group-Based Tasks" in [42]. As explained above, we selected nine representative tasks (T1 to T9), with three tasks in each category. Task descriptions and details are provided in Table 1.

## 3.2 Color Selection

Since the user study required colors to be identified by their names, we ran a pilot study to verify that the colors we use can be quickly and uniformly named by most people. This is particularly important in our case since most of the participants were not native English speakers.

We selected our colors using ColorBrewer [11]. We considered qualitative color schemes that had enough colors to cover the maximum number of data classes present in our dataset (seven), and among those we selected the one with colors that are easiest to name (see the seven colors in Figure 4). Then, we presented the colors to six participants and asked them to give a name to each color. We found a full consensus on the colors red, orange, yellow, green, blue, purple, and a slight variation on brown (called "yellowish brown" by a participant).

## 3.3 Size and Density

We chose a minimum and maximum number of nodes so that the average response time for a single task is in the range from 5 to 30 seconds. We carried out a second pilot study with six different participants to determine these values.

For two different datasets, we generated all three *Techniques* with the number of nodes ranging from 50 to 350, in increments of 50 nodes. For each of these drawings (42 in total), we asked six participants to perform the following tasks "How many nodes belong to a specific group?" and "Find node X." We measured the time required to provide an answer, obtaining times ranging from 7.3 seconds for 50 nodes, to 40.2 seconds for 350 nodes.

We finally determined $N = 50$ nodes as minimum (7.3 seconds), $4N = 200$ nodes as maximum (24.3 seconds), and $2N = 100$ nodes as an intermediate value.

Determining a good range for *Density* (number of links divided by number of nodes) is a difficult problem. We chose $L = N$ (tree-like networks) for the sparsest setting, then doubled the density to $2L$, and doubled in again to $4L$ in keeping with the geometric growth for *Size*.

## 3.4 Data

We use several real-world relational datasets for our evaluation, in order to minimize potential bias introduced by just one dataset.

- **Recipe-ingredients**, contains 350 unique cooking ingredients extracted from 50,000 cooking recipes [1]. Links are weighted based on co-occurrence of the ingredients in the recipes.

- **World-trade**, contains trade relationships between 200 countries. Links are weighted based on normalized combined import/exported between pairs of countries [21].

- **Colors**, contains 500 uniquely named colors [33] with links defined by the distance in RGB space between corresponding pairs.

The nodes in the dataset are labeled with familiar words: cooking ingredients, country names, color names. We were concerned that referring to cluster colors and node colors might be confusing (for the Colors dataset), but no participants mentioned this as a problem.

From each dataset, we selected 200, 100 and 50 nodes by iterative (random) filtering. For each dataset and each size (*Size*), we constructed a graph for each *Density* with 4, 2 and 1 times as many links as nodes, by selecting the links with highest weights.

The graphs are embedded with an MDS [28] algorithm and clustered using Modularity Clustering [35], with the link weight as similarity between connected nodes. For both algorithms, we used the implementations provided in GRAPHVIS [15].

We built GMaps [21] as instances of NLG diagrams. From there, we obtained the NL diagrams by removing the group regions, and the N diagrams by further removing the links.

## 3.5 Participants and Setting

We recruited 36 participants (23 male, 13 female) aged 21–32 years (mean 24) with normal vision. Participants were undergraduate and graduate science and engineering students, familiar with plots, graphs and networks. We divided the participants into three groups: 12 participants (8 male, 4 female) to perform tasks using N diagrams, 12 participants (7 male, 5 female) to perform tasks using NL diagrams, and 12 participants (8 male, 4 female) to perform tasks using NLG diagrams. The study was conducted on a computer with i7 CPU 860 @ 2.80GHz processor and 24 inch screen with 1600x900 pixel resolution. Participants interacted with mouse to complete the tasks.

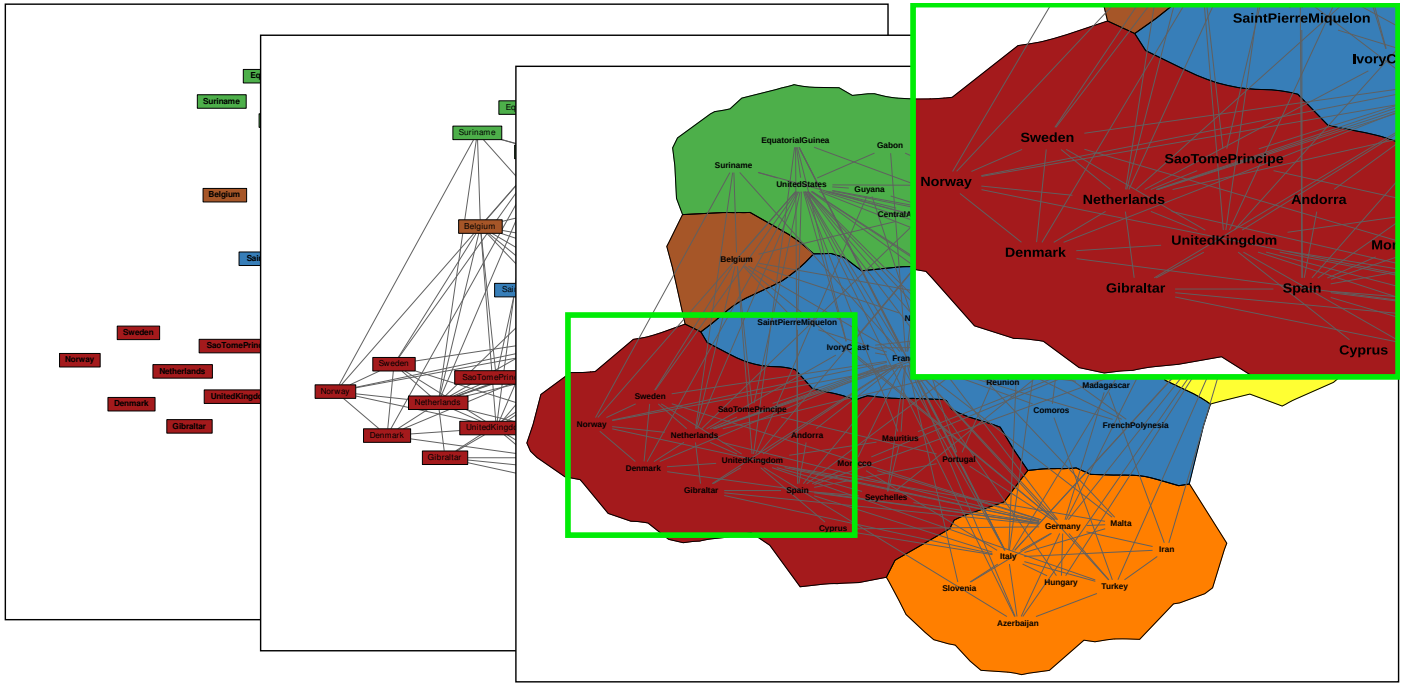

Figure 4: Representation of 50 nodes and 200 links with N, NL and NLG diagrams; underlying data from the world-trade dataset.

## 3.6 Experimental procedure

We used a full factorial between-subjects design. For each *Technique* (N, NL, NLG), we had 3 *Sizes*, 3 *Densities* and 9 *Tasks*. Each participant performed 3 *Size* × 3 *Density* × 9 *Tasks* = 81 tasks.

Before the controlled experiment, participants were briefed about the purpose of the study, data, and *Technique* used. Although all participants were familiar with graphs, we explained all technical definitions (e.g., node, links, adjacency, groups, paths). We then asked them to complete 9 training tasks as quickly and accurately as possible. The participants were encouraged to ask questions during this stage (we do not record the time and accuracy for trials).

The main experiment consisted of 81 tasks for a specific *Technique* (node N, node-link NL, or node-link-group NLG). The tasks were presented in a reduced Latin square to counterbalance learning and order effects (to prevent participants from extrapolating new judgments from previous ones). The participants were able to zoom and pan the diagram on the screen (if needed) and were required to select one of the provided multiple choices. We recorded time and accuracy for each task. The participants were instructed to take breaks if needed when they saw a blank screen. A screenshot of software for the experiment is shown in Figure 3.

## 3.7 Hypotheses

Since the three *Techniques* show information that can be either relevant or detrimental in a particular analysis scenario, we expect that each *Technique* will have its advantages and disadvantages. We collected these expectations in the following hypothesis:

- **H1:** For Node-Based tasks there will be no significant differences between the three *Techniques*, as nodes are represented in all diagrams with the same characteristics. However, NL and NLG diagrams could be penalized when the *Density* increases, since a large number of links might obstruct the detection of the nodes [25].

- **H2-a:** For Network-Based tasks, unlike in [26], we believe there will be no significant differences between NL and NLG diagrams. Although is has been shown that performance improves for map-like visualizations compared with node-link diagrams for revisitation tasks [23], we believe that for accessibility and connectivity tasks the results will be comparable, as nodes and links have the same characteristics in NL and NLG diagrams (node positions, link positions, and font size).

- **H2-b:** For Network-Based tasks, the increase of *Density* (links) and *Size* (nodes) will result in a decrease in the performances in NL and NLG diagrams.

- **H3:** Earlier work indicates no significant difference between NL and NLG diagrams for group-based tasks [26]. However, we hypothesize that for group-based tasks, NLG diagrams will outperform NL diagrams, given that the NLG diagrams have contiguous regions. We base this hypothesis on research that shows that map visualizations have two desirable features: explicit grouping and explicit group boundaries such as in [12, 21], and the observation that people tend to create layouts that distinctively group clusters in non-overlapping spatial regions [24].

## 4 Results

We first describe the methods used to analyze the data gathered from the user experiment. We then provide an overview of our results, with more detailed quantitative results listed and described in Figures 5, 6 and 7. We excluded about 26% incorrect trials for N diagrams (mostly network-based tasks), 11% for NL diagrams and 10% for NLG diagrams. Accuracy is measured using the number of correct trials divided by the total number of trials, thus showing a percentage. Time is measured in seconds.

### 4.1 Data Analysis

We evaluate the performance of different types of tasks with different *Techniques* using 2*3 between-subjects ANOVA with *Technique* (N, NL and NLG) and *Task* (node-based, network-based and group-based tasks) as factors. The main effect of *Technique* indicates which *Technique* produces the best performance, regardless of the task. The main effect of *Task* indicates which task is performed well, regardless of visualization method. The *Task* x *Technique* interaction indicated whether a particular *Technique* works better with a particular task.

In order to investigate the effects of *Density* (number of links) on user performance, we conducted 2*3 between-subjects ANOVA with *Density* (*L*, 2*L* and 4*L*) and *Task* (node-based, network-based and group-based task) as factors. We conducted this test for NL and NLG diagrams independently. (N diagrams are not considered in the *Density* analysis, as they are not affected by a change in the number of links.)

Finally, for assessing the effect of *Size* (number of nodes) on user performance, we conducted 2*3 between-subjects ANOVA with *Size* (*N*, 2*N* and 4*N*) and *Task* (node-based, network-based and group-based task) as factors. This test was performed independently for each *Technique*.

### 4.2 Result Overview

There is little change in performance of node-based tasks across the three different types of diagrams, which supports hypothesis H1. Similarly, H2 is supported by the data as network-based tasks are performed as accurately with NL as with NLG diagrams; moreover, network-based tasks are performed significantly faster with NLG diagrams than with NL diagrams. Finally, H3 is also supported by the data with statistically significant improvements in both accuracy and speed for group-based tasks using NLG diagrams, compared with NL diagrams.

Assessing the effect of a *Technique* on performance revealed that NLG diagrams are about 8% more accurate than NL diagrams, and 22% more accurate than N diagrams across all tasks. We found that tasks were performed 15% faster when using NL and NLG diagrams, compared to N diagrams, across all tasks. More details are shown in Figure 5.

*Density* affected accuracy in different ways for different tasks. Results on network-based tasks indicates significant difference in accuracy (when comparing *Density L* with *Density* 4*L*) for NL and NLG diagrams. However, for node-based tasks and group-based tasks, despite a slightly decreased accuracy with increased Density, there were no statistically significant differences. *Density* affected time performance differently as well. Both node-based and network-based tasks were significantly slower (when comparing *Density L* with *Density* 4*L*) for NL and NLG diagrams. However, node-based tasks again were mostly unaffected. More details are shown in Figure 6.

*Size* affected accuracy only for network-based tasks. More specifically, network-based tasks show significant decrease in accuracy with NL and NLG diagrams and only when the *Size* is quadrupled (when comparing *Size N* with *Size* 4*N*). *Size* had much greater impact on time performance across all types of tasks and all types of diagrams. Node-based tasks were significantly slower (when comparing *Density L* with *Density* 4*L*) for N, NL, and NLG diagrams. Network-based tasks were significantly slower (when comparing *Density L* with *Density* 4*L*) for NL, and NLG diagrams. Group-based tasks were significantly slower (when comparing *Density L* with *Density* 4*L*) only for NL diagrams. More details are shown in Figure 7.

## 5 Discussion

In our experiments we attempted to control several variables that typically impact such studies. In particular, for a given dataset, we fixed the location of the nodes in the N, NL, and NLG diagrams. We also fixed the links in the NL and NLG diagrams. We also used the same font size and the same colors to indicate groups in all diagrams. This allows us to focus on the impact of varying *Size* and *Density*, across diagrams.

There is little change in performance of node-based tasks across the three different types of diagrams, which supports hypothesis H1; see Fig. 5. Moreover, we note that high link *Density* penalizes node-based task time performance in both NL and NLG diagrams, confirming the second part of H1; see Fig. 6(c,d). We believe that this happens because links are only a distraction for node-based tasks, but their negative effect is mitigated by the fact that they are drawn behind the nodes in our drawings.

Similarly, H2-a is supported by the data as network-based tasks are performed as accurately with NL as with NLG diagrams; see Fig. 5(a); moreover, network-based tasks are performed significantly faster with NLG diagrams than with NL diagrams; see Fig. 5(b). This contradicts results in Jianu et al. [26], where a NL diagrams (called "node coloring") performed better than NLG diagrams (GMap and BubbleSets) for network-based tasks. We believe that this is due to the absence of fragmentation and the better choice of colors in our maps, as well as the absence of group labels (which obscured important information in their experiments).

The increase of *Density* and *Size* result in a decrease in the network-task performance (both time and accuracy) supporting H2-b; see Fig. 6-7. This confirms results in the study of Alper et al. [2]. However, our initial expectations of a drastic performance reduction for the maximum link *Density* were not confirmed by our experiment, most likely because our maximum parameters were not very high (quadrupling the *Size* and *Density*).

Finally, H3 is also supported by the data with statistically significant improvements in both accuracy and speed for group-based tasks using NLG diagrams, compared with NL diagrams; see Fig. 5. This contradicts the results in Jianu et al. [26], but is likely explained with use of fragmented maps in their study and contiguous maps in ours.

Finally *Size* and *Density* do not influence the performances of group-based tasks in all diagrams (except there is a negative effect of *Size* on the performance time of group-based tasks in NL diagrams); see Fig. 7(e). This is likely due to the fact that the number of clusters remains constant when *Density* and *Size* increase.

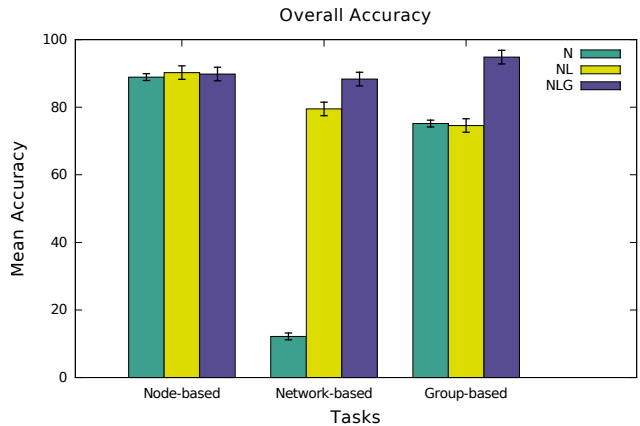 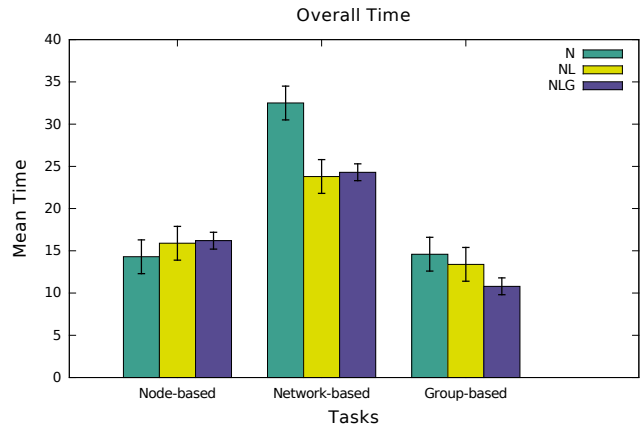

Figure 5: (a) Mean accuracy (in percentage) for three different categories of tasks in different diagrams, (b) Mean completion time (in seconds) for three different categories of tasks in different diagrams. Error bars represent +/-2 standard deviation.

While the variations between tasks in the same category were generally not very large, T9 in the third category was an exception. Specifically, that average performance time and accuracy for T9 with N and NL diagrams are significantly worse than with NLG diagrams. This is the main reason why NLG diagrams outperform N and NL diagrams in the group-based category. We believe that this could be explained with the explicit presence of boundaries for the groups in NLG diagrams, which are absent in N and NL diagrams.

One of the main findings in our study is that NLG diagrams perform well across all tasks. While it is not very surprising that NLG diagrams perform well for group-based tasks, it is somewhat unexpected that NLG diagrams outperform NL diagrams, and offer the same performance for node-based tasks, in our setting.

# 6  Conclusions and Future Work

We provide online (https://sites.google.com/site/infovispaper) all relevant materials for this study: the three datasets used in our experiment (colors graph, world-trade graph, and recipe-ingredients graph), the software for running the experiment, and the results (accuracy and time) of the 2,916 individual trials. We consider this the first in a series of controlled studies to evaluate the advantages and disadvantages of node, node-link, and node-link-diagram visualizations. In this experiment we considered the impact of *Size* and *Density* on standard node, network, and group tasks using the three visualizations. We did not address more sophisticated issues, such as knowledge discovery, knowledge retention, engagement, enjoyment-factors, intimidation-factors, and interaction. We anticipate that NLG diagrams will outperform N and NL diagrams, but this remains to be studied.

The good performance of NLG diagrams in our study suggests that more work is needed to evaluate different NLG diagram generation methods such as Bubblesets, Linesets, Kelp diagrams, and GMap.

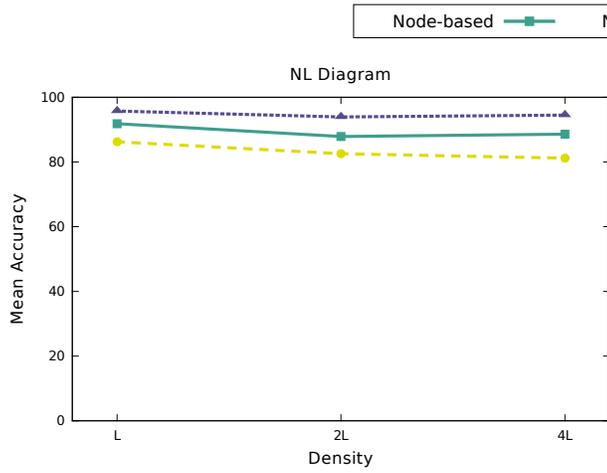

**Significance**
Density($F_{(2,99)} = 27.1, p < .05$)
Task ($F_{(2,99)} = 20.4, p < .05$)
Task x Density ($F_{(4,99)} = 9.52, p < .001$)

**Pairwise Comparisons (Posthoc Tukey's HSD)**
Network-based tasks    Density L vs. 4L    ($p < .05$)

**Result Explanation**
Accuracy for network-based tasks in NL diagram significantly decreases (from 86.4% to 75.5%) when *Density* quadruples. Increasing *Density* does not have significant effect on accuracy of node-based and group-based tasks in NL diagrams.

(a)

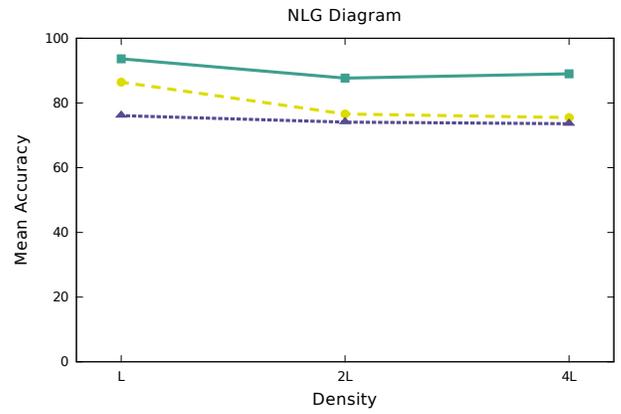

**Significance**
Density($F_{(2,99)} = 40.3, p < .001$)
Task ($F_{(2,99)} = 23.2, p < .001$)
Task x Density ($F_{(4,99)} = 16.01, p < .001$)

**Pairwise Comparisons (Posthoc Tukey's HSD)**
Network-based tasks    Density L vs. 4L    ($p < .001$)

**Result Explanation**
Accuracy of network-based tasks significantly decreases (from 85.24% to 80.17%) when *Density* quadruples in NLG diagrams. However, changes in *Density* does not have significant effect on accuracy for node-based and group-based tasks in NLG diagrams.

(b)

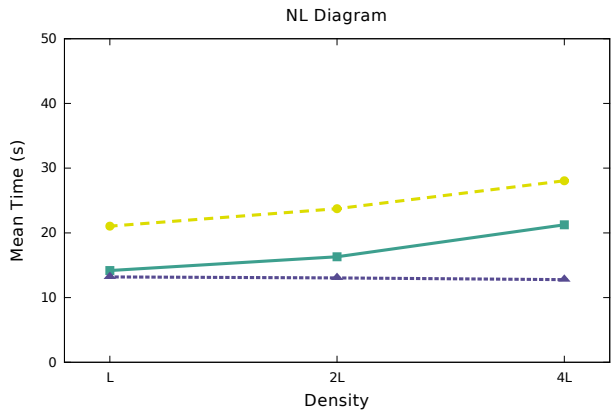

**Significance**
Density($F_{(2,99)} = 27.2, p < .05$)
Task ($F_{(2,99)} = 30.1, p < .001$)
Task x Density ($F_{(4,99)} = 12.5, p < .001$)

**Pairwise Comparisons (Posthoc Tukey's HSD)**
Node-based tasks       Density L vs. 4L    ($p < .001$)
Network-based tasks    Density L vs. 4L    ($p < .05$)

**Result Explanation**
Time for node-based and network-based tasks significantly increases (33% and 25%) when *Density* quadruples in NL diagrams. Changes in *Density* do not have a significant effect on time for group-based tasks in NL diagrams.

(c)

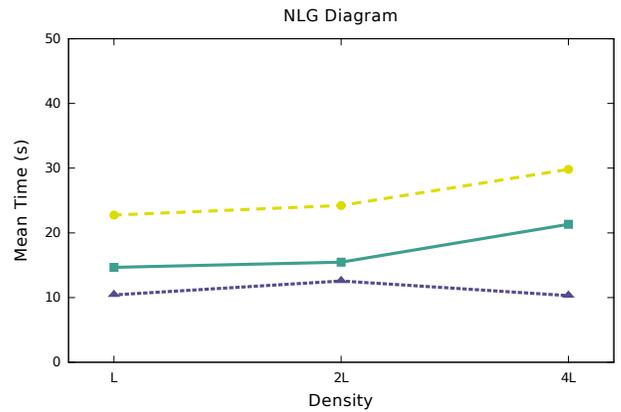

**Significance**
Density($F_{(2,99)} = 22.83, p < .001$)
Task ($F_{(2,99)} = 14.7, p < .001$)
Task x Density ($F_{(4,99)} = 10.4, p < .05$)

**Pairwise Comparisons (Posthoc Tukey's HSD)**
Node-based tasks       Density L vs. 4L    ($p < .001$)
Network-based tasks    Density L vs. 4L    ($p < .001$)

**Result Explanation**
Time for node-based and network-based tasks performance time significantly increases (30% and 22%) when *Density* quadruples. However, changes in *Density* do not have a significant effect on time for group-based performance in NLG diagrams.

(d)

Figure 6: Performance and Time accuracy for three different categories of tasks with different densities (*L*, 2*L* and 4*L*). **Top:** Mean completion time (in seconds) for three different categories of tasks for NL and NLG diagrams, **Bottom:** Mean accuracy (in percentage) for three different categories of tasks. We excluded N diagram from *Density* analysis since changes in number of links/edges does not have any effect on N diagram.

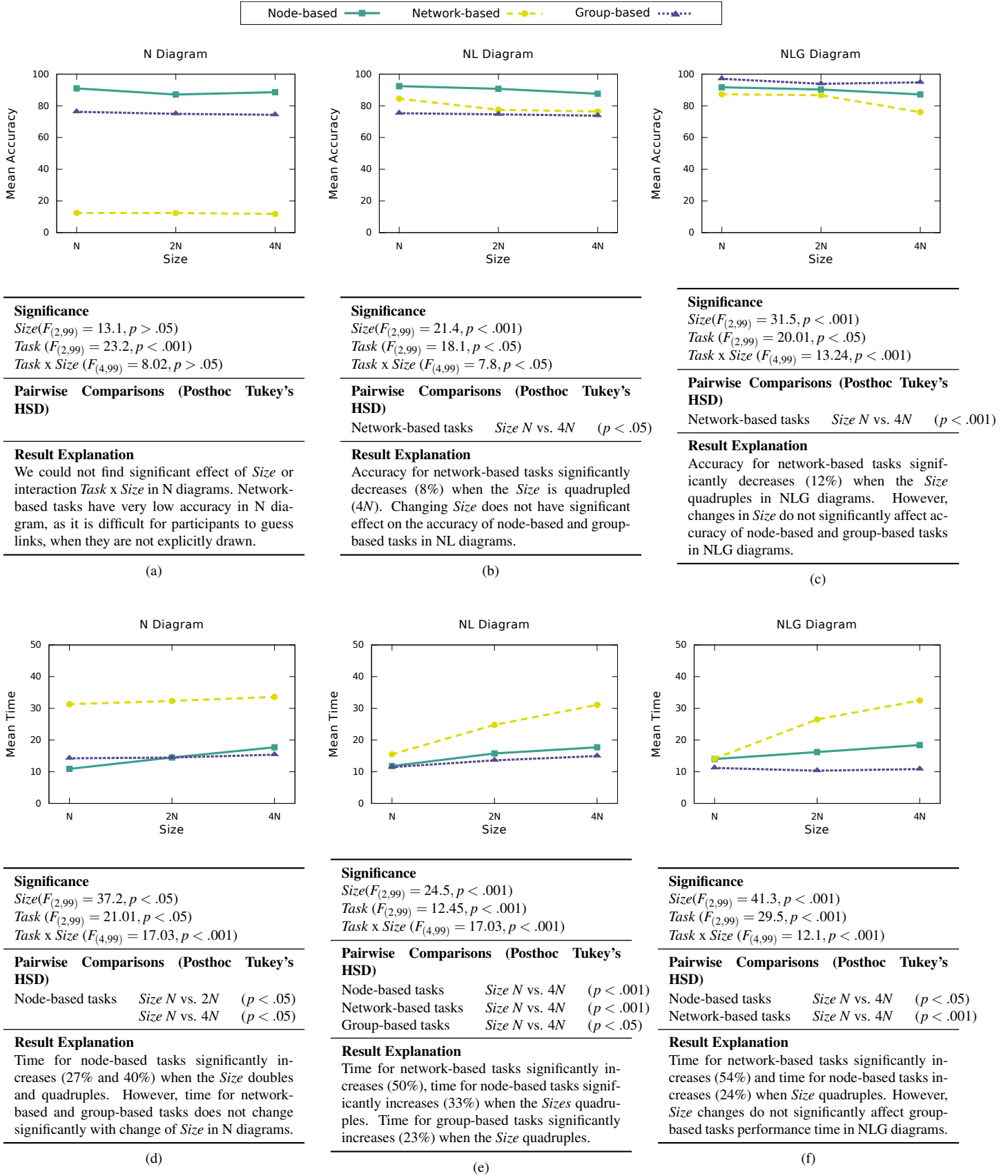

Figure 7: Performance and Time accuracy for three different categories of tasks with different *Sizes* (*N*, 2*N* and 4*N*). **Top:** Mean completion time (in seconds) for three different categories of tasks for N, NL and NLG diagrams, **Bottom:** Mean accuracy (in percentage) for three different categories of tasks.